# Bounds on Attainable Rate Modulation in Fermi's Golden Rule

Kaisei Ohata[1], Jun-Yu Dong[1], Yasutaka Kitahama[1,2,3]*, Akihito Ishizaki[1], and Keisuke Goda[1,2,3,4,5,6,7,8,9]*

[1]*Department of Chemistry, The University of Tokyo, Tokyo 113-0033, Japan*

[2]*LucasLand, Tokyo 101-0052, Japan*

[3]*Institute for Quantum Life Science, National Institutes for Quantum Science and Technology, Chiba 263-8555, Japan*

[4]*Graduate School of Medicine, Tohoku University, Miyagi 980-0872, Japan*

[5]*SiRIUS Institute of Medical Research, Tohoku University, Miyagi 980-0872, Japan*

[6]*International Center for Synchrotron Radiation Innovation Smart, Tohoku University, Miyagi 980-8577, Japan*

[7]*Healthspan Research Center, Tohoku University, Miyagi 980-0872, Japan*

[8]*Institute of Technological Sciences, Wuhan University, Hubei 430072, China*

[9]*Department of Bioengineering, University of California, Los Angeles, CA 90095, USA*

* Contact authors: kitahama@g.ecc.u-tokyo.ac.jp, goda@chem.s.u-tokyo.ac.jp

**ABSTRACT**. Controlling transition rates is a central objective across quantum optics, condensed-matter physics, and chemical physics, encompassing processes such as spontaneous emission, energy and charge transfer, spin relaxation, and chemical reactions. Although Fermi's golden rule provides a powerful framework for calculating such rates of weakly coupled transitions, conventional calculations require the interaction to be specified in advance; the range of rate responses attainable over a broader interaction space remains unknown. We formulate spectral-overlap golden-rule rates as quadratic forms of the interaction and derive rigorous bounds on rate modulation over a specified interaction space. As an example, application to energy transfer from a donor to a molecular vibration coupled to an optical cavity reveals that enhancement and suppression depend jointly on the interaction and cavity detuning, including continuous reversal of the rate response for a fixed interaction. The same approach may elucidate rate control for other golden-rule processes, including excitation-energy transfer, nonadiabatic electron transfer, nonradiative electronic transitions, vibrational and spin relaxation, and spontaneous emission.

*Introduction.*—Controlling transition rates is a central objective across quantum optics, condensed-matter physics, and chemical physics, encompassing processes such as spontaneous emission, energy and charge transfer, spin relaxation, and chemical reactions [1–4]. Such control underlies broader goals ranging from the generation and preservation of quantum states to efficient energy conversion and selective chemical transformations [5–8]. For example, suppressing relaxation in solid-state spin qubits can extend spin-state lifetimes and reduce a fundamental limitation on the time available for coherent quantum-information processing [9,10]. Coupling molecular vibrational or electronic degrees of freedom to confined electromagnetic modes also provides a means of modifying transition rates [11–13]. Such cavity-induced modifications are being explored to control energy flow, chemical reactivity, and reaction selectivity [14–16]. Predicting the magnitude and direction of such rate changes is therefore essential for assessing the extent of achievable dynamical control.

For weakly coupled transitions, Fermi's golden rule provides a broadly applicable framework for calculating rates from the available states and the interaction connecting them. Its ability to combine energetic matching with interaction matrix elements underlies its use across a wide range of rate processes [17–20]. However, most such calculations are naturally formulated as forward problems: the interaction driving a transition is specified first, and the corresponding rate is then evaluated. In nonadiabatic electron transfer, for example, the Condon approximation treats the electronic coupling as independent of nuclear coordinates, whereas non-Condon formulations retain coordinate-dependent couplings [21,22]. Although the latter admits a broader class of interactions, a particular interaction must still be specified before the rate is evaluated. Such forward calculations predict a specified process but pose a limitation for rate-control design: a rate change calculated for one interaction form does not reveal how strongly the result depends on that choice or what responses are possible over a broader class of interactions. Without knowing the limits of these possible responses, one cannot determine whether a desired degree of rate control is possible within a given interaction space or whether an apparently weak response leaves room for substantially greater modulation.

In this Letter, we reverse the usual order of the calculation. Rather than choosing a particular interaction and then calculating its rate, we first ask what range of rate changes is possible over a specified interaction space. We show that this range can be obtained by retaining the dependence of the golden-rule rate on the interaction, yielding rigorous upper and lower bounds. Individual interactions can then be examined to determine how responses within those bounds are realized. We demonstrate the approach for energy transfer to an anharmonic molecular vibration coupled to a cavity. The bounds establish the range of rate changes accessible at each cavity condition, whereas a subsequent search resolves how individual interactions respond as the cavity is detuned. We identify a fixed interaction whose response can be switched continuously from enhancement through cancellation to suppression by tuning only the cavity frequency, showing that the magnitude and direction of cavity-induced rate modulation are determined jointly by the cavity-modified system and the microscopic interaction driving the process. The framework is not specific to cavity-modified systems and applies more generally to determining the range of spectral-overlap golden-rule rate changes between different physical conditions.

*Formulation.*—We consider a transition into a quantum system described by a Hamiltonian $H$, with eigenstates and energies

$$H|\lambda\rangle = E_\lambda|\lambda\rangle.$$

The system is initially in an eigenstate $|\mathrm{I}\rangle$, with $H|\mathrm{I}\rangle = E_\mathrm{I}|\mathrm{I}\rangle$, and the transition is driven by an interaction operator $X$. We consider spectral-overlap Fermi golden rule rates of the form

$$k_H[X] = C \sum_{\lambda(\neq \mathrm{I})} S(E_\lambda - E_\mathrm{I})|\langle\lambda|X|\mathrm{I}\rangle|^2, \tag{1}$$

where $C$ collects factors independent of the Hamiltonian and interaction operator, and $S(E)$ weights transitions according to their energy gaps.

Equation (1) gives the rate once a particular interaction $X$ has been specified. To retain the dependence on the interaction before evaluating the scalar rate, we restrict $X$ to a specified interaction space with operator basis $\{O_m\}_{m=1}^{M}$,

$$X(\mathbf{x}) = \sum\nolimits_{m=1}^{M} x_m O_m, \tag{2}$$

where the coefficients $x_m$ may in general be complex. Substitution into Eq. (1) gives

$$k_H[\mathbf{x}] = C\mathbf{x}^\dagger K_H \mathbf{x}, \tag{3}$$

where the rate kernel is

$$(K_H)_{mn} = \sum_{\lambda(\neq \mathrm{I})} S(E_\lambda - E_\mathrm{I})\langle \mathrm{I}|O_m^\dagger|\lambda\rangle\langle\lambda|O_n|\mathrm{I}\rangle. \tag{4}$$

The matrix $K_H$ is Hermitian and positive semidefinite, whereas $\mathbf{x}$ selects a particular interaction within the chosen interaction space.

The quadratic dependence of transition observables on a set of coefficients has previously been exploited in several optimization settings. Examples include optimizing transition rates, yields, and related observables over dipole orientations or coherent superpositions, as well as optimizing selectivity through ratios of quadratic observables [23–26]. We use this structure to determine the range of relative rate modulation between a reference system $H_0$ and a modified system $H_1$. The corresponding rate kernels $K_j$ are constructed from the eigenstates, energies, and initial state $|\mathrm{I}_j\rangle$ of each system, whereas the same interaction $X(\mathbf{x})$ is evaluated in both.

Defining $\Delta K = K_1 - K_0$, the relative modulation becomes

$$\eta[\mathbf{x}] = \frac{k_1[\mathbf{x}] - k_0[\mathbf{x}]}{k_0[\mathbf{x}]} = \frac{\mathbf{x}^\dagger \Delta K \mathbf{x}}{\mathbf{x}^\dagger K_0 \mathbf{x}}. \tag{5}$$

Because Eq. (5) is invariant under an overall rescaling of $\mathbf{x}$, the relative modulation depends only on the interaction direction within the chosen interaction space. It is a generalized Rayleigh quotient whose stationary values satisfy

$$\Delta K \mathbf{x} = \eta K_0 \mathbf{x}. \tag{6}$$

On the subspace where $K_0$ is positive definite, the generalized Rayleigh quotient obeys

$$\eta_\mathrm{min} \leq \eta[\mathbf{x}] \leq \eta_\mathrm{max},$$

where $\eta_\mathrm{min}$ and $\eta_\mathrm{max}$ are the smallest and largest generalized eigenvalues of Eq. (6). These are global extrema over the selected interaction space, whereas the corresponding generalized eigenvectors identify the interaction directions that attain the limiting responses. Interactions for which $\mathbf{x}^\dagger K_0 \mathbf{x} = 0$ do not define a finite relative modulation and are thus excluded; numerical details and convergence tests are provided in the Supplemental Material [41].

The construction therefore separates two questions that are combined in a conventional forward rate calculation. The bounds first determine what range of relative rate changes between the reference and modified systems is possible over a specified interaction space. We then examine individual interactions to determine which responses they realize and how those responses evolve as the system is varied.

*Model.*—We apply the framework to energy transfer from an external donor to an anharmonic molecular vibration coupled to a cavity. The donor is represented by its spectral distribution, whereas the acceptor vibration and its cavity-induced modification are treated explicitly through the molecular–cavity Hamiltonian. We assume sufficiently weak donor–acceptor coupling for the transfer to be described by a spectral-overlap Fermi-golden-rule rate [17]. A possible physical realization is an electronically excited donor transferring energy to vibrational states of an acceptor. The cavity can then couple predominantly to the acceptor vibration, whereas the donor remains outside the strongly coupled subsystem and enters the rate through its spectrum.

We describe the acceptor by an anharmonic vibrational Hamiltonian $H_{\mathrm{vib}}$, with eigenstates $|\nu\rangle$ and energies $\varepsilon_\nu$,

$$H_{\mathrm{vib}}|\nu\rangle = \varepsilon_\nu|\nu\rangle. \quad (7)$$

We introduce a dimensionless vibrational coordinate $Q$, normalized such that

$$\langle 0|Q|0\rangle = 0, \qquad \langle 1|Q|0\rangle = 1.$$

This normalization provides a common dimensionless scale for the light–matter coupling and coordinate-dependent interactions.

The coupled molecule–cavity system is described by the single-mode dipole-gauge Pauli–Fierz Hamiltonian [27]

$$H_1 = H_{\mathrm{vib}} + \omega_{\mathrm{C}} a^\dagger a + gQ(a + a^\dagger) + \frac{g^2}{\omega_{\mathrm{C}}} Q^2. \quad (8)$$

Here, $a^\dagger$ and $a$ create and annihilate a cavity photon of frequency $\omega_{\mathrm{C}}$, $g$ is the light–matter coupling strength, and the final term is the dipole self-energy. Equation (8) retains both rotating and counterrotating interaction terms. The uncoupled reference is

$$H_0 = H_{\mathrm{vib}} + \omega_{\mathrm{C}} a^\dagger a. \quad (9)$$

The coupled and reference Hamiltonians are represented in the same truncated vibrational–photon basis at each cavity frequency. We write

$$\omega_{\mathrm{C}} = f\omega_{01}, \qquad \omega_{01} = \varepsilon_1 - \varepsilon_0,$$

and vary $f$ to map the cavity-detuning dependence. For the present model, the initial states $|\mathrm{I}_0\rangle$ and $|\mathrm{I}_1\rangle$ are taken to be the ground states of $H_0$ and $H_1$, respectively. For each $f$, we diagonalize Eqs. (8) and (9) independently and construct the corresponding spectral operators and rate kernels.

For the numerical demonstration, we take $H_{\mathrm{vib}}$ to be a Morse oscillator and use $g = 0.1\omega_{01}$, with the cavity frequency varied over $0.8 \leq f \leq 1.2$. The donor is represented by a Gaussian spectrum centered at $\Omega_D = 3\omega_{01}$ with standard deviation $\sigma_D = 0.20\omega_{01}$, placing the transfer in an energy range where transitions to higher-lying anharmonic vibrational states of the acceptor contribute. The transitions entering Eq. (1) are assigned a Lorentzian half-width at half-maximum $\gamma_A = 0.08\omega_{01}$. The convolution of the Gaussian donor spectrum with the Lorentzian transition lines gives the Voigt spectral-overlap function $S(E_\lambda - E_I)$ entering Eq. (1). Details of the Morse oscillator used in the numerical calculations, the coordinate normalization, and numerical convergence are provided in the Supplemental Material [41].

*Results.*—As a minimal physically motivated interaction space, we consider the operator basis $\{Q, Q^2, Q^3\}$, so that

$$X(\mathbf{x}) = x_1 Q + x_2 Q^2 + x_3 Q^3, \mathbf{x} = (x_1, x_2, x_3)^\intercal. \quad (10)$$

Such a low-order expansion arises naturally when the physical quantity governing the transfer depends smoothly on a vibrational coordinate and is expanded about a reference value of that coordinate. Coordinate-dependent transition moments or vibronic couplings provide familiar examples [28,29]. We use $Q$, $Q^2$, $Q^3$ as a minimal operator basis spanning linear and nonlinear coordinate dependences, rather than as three distinct microscopic mechanisms.

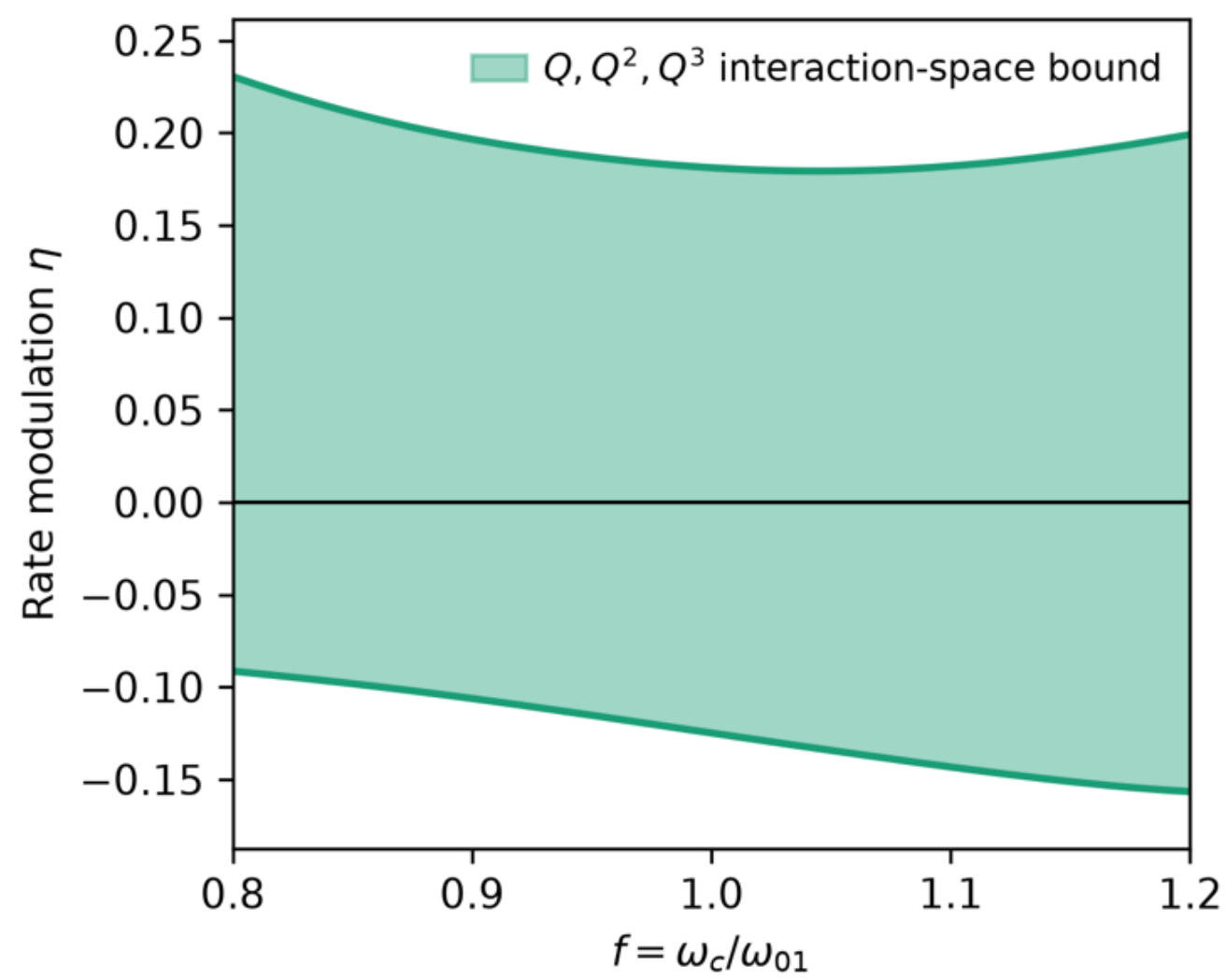

FIG. 1. Bounds on cavity-induced rate modulation within the $Q$, $Q^2$, $Q^3$ interaction space. The lower and upper curves show the minimum and maximum relative rate modulations $\eta$ obtained from the generalized eigenvalue problem as a function of the normalized cavity frequency $f = \omega_{\mathrm{C}}/\omega_{01}$. The shaded region between the bounds represents the full range of rate modulations accessible within the interaction space at each cavity frequency. Positive and negative values denote enhancement and suppression, respectively.

Within this interaction space, the lower bounds range from $-15.7\%$ to $-9.15\%$, and the upper bounds from $+17.9\%$ to $+23.0\%$ over the detuning interval [Fig. 1]. Because the lower bound remains negative and the upper bound positive throughout this interval, both enhancement and suppression are accessible within the interaction space at every cavity frequency examined.

We next examine how the rate modulation varies among individual interactions within the interaction space. We fix the linear coefficient in Eq. (10) and write

$$X_{\alpha\beta} = Q + \alpha Q^2 + \beta Q^3, \qquad \mathbf{x}_{\alpha\beta} = (1, \alpha, \beta)^\intercal. \quad (11)$$

The corresponding modulation is

$$\eta(\alpha, \beta; f) = \frac{\mathbf{x}_{\alpha\beta}^\dagger \Delta K(f) \mathbf{x}_{\alpha\beta}}{\mathbf{x}_{\alpha\beta}^\dagger K_0(f) \mathbf{x}_{\alpha\beta}}.$$

Figures 2(a), 2(b), and 2(c) show $\eta(\alpha, \beta; f)$ at $f = 0.8$, 1.0, and 1.2, respectively. For visualization, we restrict the interaction family to avoid large nonlinear excursions. We retain interactions whose magnitude remains comparable to that of $Q$ over the region where the bare vibrational ground-state probability is appreciable,

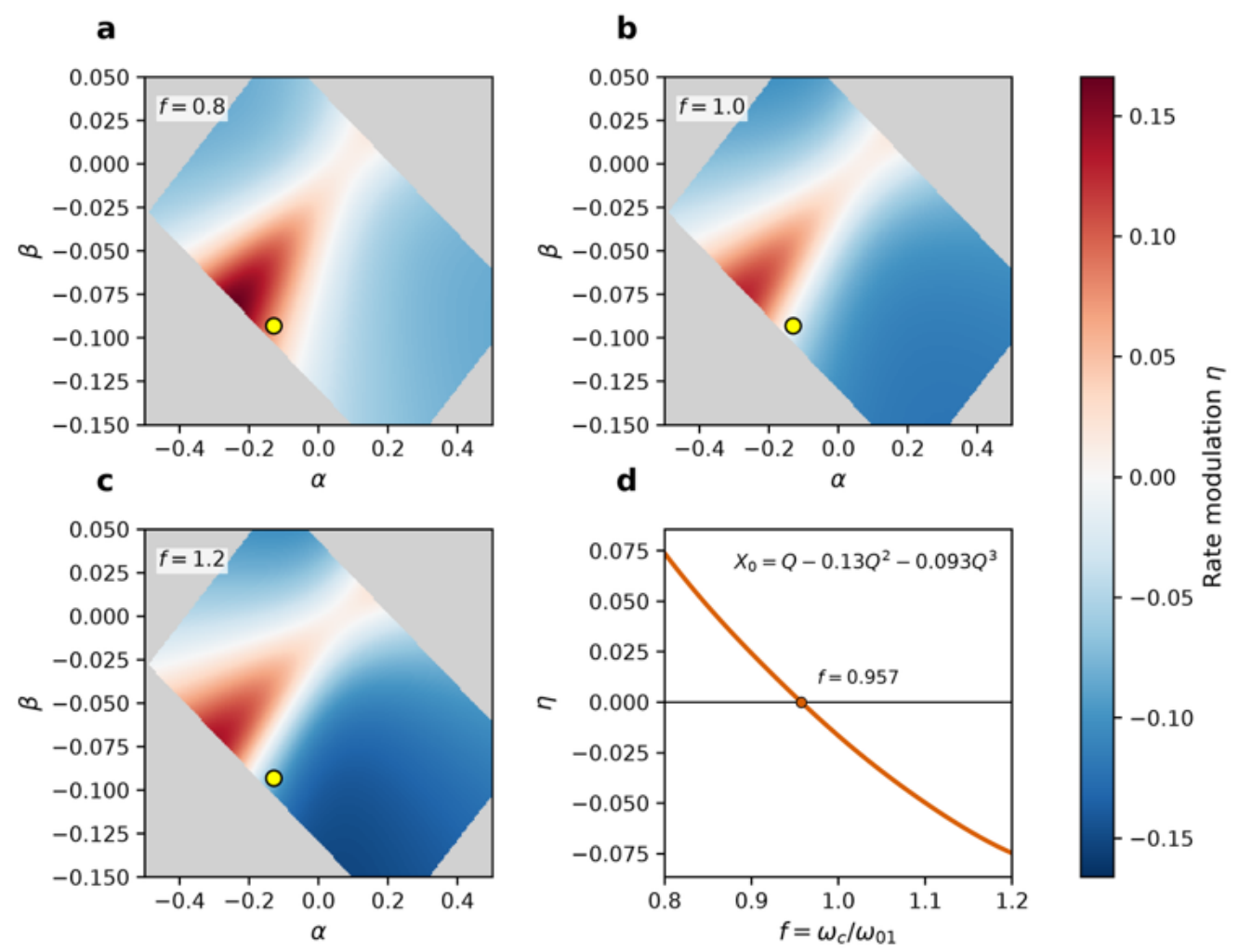

FIG. 2. Dependence of rate modulation on interaction and cavity detuning. (a)–(c) Relative rate modulation $\eta(\alpha, \beta; f)$ for $X_{\alpha\beta} = Q + \alpha Q^2 + \beta Q^3$ at $f = 0.8, 1.0,$ and $1.2$, respectively. Red and blue indicate enhancement and suppression, respectively, while white indicates values near cancellation ($\eta \approx 0$). Gray regions correspond to interactions excluded by the criterion $R_X < 2$. The yellow marker denotes $(\alpha, \beta) = (-0.13, -0.093)$. (d) Detuning dependence of the fixed interaction marked in (a)–(c), $X_0 = Q - 0.13Q^2 - 0.093Q^3$. The horizontal line denotes $\eta = 0$, which $X_0$ crosses at $f = 0.957$ as its response changes continuously from enhancement to suppression.

$$R_X(\alpha, \beta) = \frac{\max_{Q\in\mathcal{S}} |Q + \alpha Q^2 + \beta Q^3|}{\max_{Q\in\mathcal{S}} |Q|} < 2,$$

where $\mathcal{S}$ is the region in which the ground-state probability density exceeds $10^{-4}$ of its maximum. The sensitivity of the retained interaction region to the cutoff in $R_X$ and to the definition of $\mathcal{S}$ is examined in the Supplemental Material [41]. At $f = 0.8$, the retained interactions span rate modulations from $-8.4\%$ to $+16.6\%$. The corresponding intervals at $f = 1.0$ and 1.2 are $-12.1\%$ to $+12.8\%$ and $-15.2\%$ to $+13.7\%$, respectively.

The $\eta = 0$ boundary also shifts across the $(\alpha, \beta)$ plane as the cavity is detuned, showing that detuning changes not only the magnitude of the rate response but also which interactions produce enhancement or suppression. Consequently, a fixed interaction can change the sign of its response as the cavity is detuned. This behavior is illustrated by

$$X_0 = Q - 0.13Q^2 - 0.093Q^3,$$

for which the rate modulation changes from $+7.38\%$ at $f = 0.8$ to $-7.48\%$ at $f = 1.2$, crossing $\eta = 0$ at $f = 0.957$ [Fig. 2(d)], suggesting that cavity detuning can continuously switch the response of a fixed interaction from enhancement through cancellation to suppression.

*Conclusion.*—We have formulated spectral-overlap Fermi-golden-rule rates as quadratic forms of the interaction and used this structure to determine the full range of relative rate changes accessible over a specified interaction space. Applying the framework to energy transfer into an anharmonic cavity-coupled vibration, we find that enhancement and suppression are both accessible at every cavity frequency examined, whereas a fixed interaction can switch continuously from enhancement through cancellation to suppression as the cavity is detuned.

These results provide a useful perspective on the diverse rate responses reported for cavity-coupled molecular systems, including enhanced, suppressed, and weakly modified dynamical pathways under vibrational strong coupling [30–33]. Enhancement, suppression, or no change in the rate does not characterize the effect of strong coupling alone; the response depends jointly on the cavity-modified spectral structure and the microscopic interaction driving the process. For example, a weak or vanishing rate change for a particular interaction does not preclude substantial enhancement and suppression for other interactions within the same cavity-modified system. The bounds distinguish these possibilities by separating the range of responses accessible within a given interaction space from the response realized by a particular microscopic interaction.

More broadly, detailed descriptions of microscopic interactions can require multiple independent components. Coordinate-dependent interactions, for example, may involve nonlinear terms beyond constant or linear approximations [34,35]. In such cases, the present formulation allows this flexibility to be retained while determining the full range of rate modulation over all combinations of these components, rather than drawing conclusions from selected coefficient values. For rate-control design, these bounds further quantify how much modulation a given system and physical condition can support, providing a criterion for whether a target response is attainable or whether a different system or condition must be sought. Moreover, the formulation is not specific to cavity-modified systems. It applies whenever a weak-coupling transition rate can be expressed as a quadratic function of the interaction weighted by the spectral structure of the initial and final states. This includes familiar golden-rule descriptions of excitation-energy transfer, nonadiabatic electron transfer, nonradiative electronic transitions, vibrational and spin relaxation, and weak-coupling spontaneous emission [36–40].

This work was supported by the MEXT Quantum Leap Flagship Program (JPMXS0120330644) and JST ASPIRE Program (JPMJAP2316).